\documentclass[%
 reprint, floatfix,
 amsmath,amssymb,
 aps,
]{revtex4-1}

\usepackage{graphicx}
\usepackage{dcolumn}
\usepackage{bm}

\usepackage{subfigure}
\usepackage{color}

\graphicspath{{figs/}}

\begin{document}

\preprint{APS/123-QED}

\title{Actin Networks Voltage Circuits}

\author{Stefano Siccardi}
\author{Andrew Adamatzky}
\affiliation{Unconventional Computing Laboratory, Department of Computer Science, University of the West of England, Bristol, UK} 
\author{Jack Tuszy\'{n}ski}
\affiliation{Department of Oncology, University of Alberta, Edmonton, AB T6G 1Z2, Canada;
DIMEAS, Politecnico di Torino, Corso Duca degli Abruzzi 24, 10129, TO, Turin, Italy} 
\author{Florian Huber}
\affiliation{Netherlands eScience Center, Science Park 140, 1098 XG Amsterdam, The Netherlands} 
\author{J{\"o}rg Schnau{\ss}}
\affiliation{Soft Matter Physics Division, Peter Debye Institute for Soft Matter Physics, Faculty of Physics and Earth Science, Leipzig University, Germany \& Fraunhofer Institute for Cell Therapy and Immunology (IZI), DNA Nanodevices Group, Leipzig, Germany}

%

\date{\today}

\begin{abstract}
\noindent
Starting with an experimentally observed networks of actin bundles, we model their network structure in terms of edges and nodes. We then compute and discuss the main electrical parameters, considering the bundles as electrical wires. A set of equations describing the network is solved with several initial conditions. Input voltages, that can be considered as information bits, are applied in a set of points and output voltages are computed in another set of positions. We consider both an idealized situation, where point-like electrodes can be inserted in any points of the bundles and a more realistic one, where electrodes lay on a surface and have typical dimensions available in the industry.  We find that in both cases such a system can implement the main logical gates and a finite state machine.
\end{abstract}

\maketitle

\section{Introduction}

Actin filaments (AFs) and tubulin microtubules (MTs) represent the key components of cytoskeleton networks~\citep{huber2013advances}. They have been experimentally demonstrated and modelled as ionic wave conducting biowires~\cite{tuszynski2004ionic,priel2006ionic,sataric2009nonlinear,sataric2010solitonic,priel2008nonlinear,sekulic2011nonlinear,sataric2011ionic}, and predicted to support conformational solitons~\cite{hameroff1982information,hameroff2002conduction,hagan2002quantum}, as well as orientational transitions of dipole moments~\cite{tuszynski1995ferroelectric,brown1997dipole,cifra2010electric}. These propagating localizations could carry information and transform it when interacting with one another. Hence, these cytoskeleton networks could be used as nano-scale computing devices~\cite{adamatzky2018towards}. This idea dates back to an early concept of subcellular computing on cytoskeleton networks~\cite{hameroff1989information,rasmussen1990computational,hameroff1990microtubule} and was later developed further in the context of information processing on actin-tubulin networks of neuron dendrites~\cite{priel2006dendritic}.
When immersed in an ion-rich liquid environment, AFs can be viewed as wires that can conduct electrical currents~\cite{tuszynski2004ionic,tusz2018}. Previously we have demonstrated computationally that these electrical currents can be used to implement Boolean gates~\cite{siccardi2016boolean}. A single AF hence can be conceived as a computing device in computational experiments. Its practical implementation under laboratory conditions, however, would be very challenging and likely beyond current technological possibilities. For this reason, we decided to adapt our computing schemes to more realistic scenarios of bundles of AFs instead of single AF units~\cite{schnaussreview2016,schnaussPRL2016}. We developed a model of an actin droplet computer, where information is represented by travelling spikes of excitation and logical operations are implemented at the junctions of AF bundles~\cite{AdamatzkyBundles2019,adamatzky2019actin,huber2015formation}. The model developed treats the actin network as a continuum with propagating abstract excitation waves -- modelled with FitzHugh-Nagumo equations. The model might be phenomenologically correct, but is not able to sufficiently describe the physics of the waves in the AF networks. Therefore, we here propose a model more solidly rooted in the underlying physics. We consider the AF networks to be made of wires and their bundles to be connected at node locations. Each bundle has its own set of electrical parameters and facilitates the movement of ions along its length. In the following we present the model in full detail and discuss the results derived from the model.

\section{The model}

A detailed description of key models that we will use as foundation in this study, can be found in \cite{tuszynski2004ionic}, which was aiming at a description of AFs. Let us highlight the assumptions on which the model was build. Each monomer in the filament has 11 negative excess charges. The double helical structure of the filament provides regions of uneven charge distribution such that pockets of higher and lower charge density exist. There is a well-defined distance, the so-called Bjerrum length $\lambda_{B}$, beyond which thermal fluctuations are stronger than the electrostatic attraction or repulsion between charges in solution. It is inversely proportional to temperature and directly proportional to the ions' valence z:
\begin{equation}
\lambda_{B} = \frac{z e^2}{4 \pi \epsilon \epsilon_0 k_B T},
\label{bjerrum}
\end{equation}
where $e$ is the electrical charge, $\epsilon_0$ the permittivity of the vacuum, $\epsilon$ the dielectric
constant of the solution with AFs immersed in (estimated similar to $\epsilon_{water} \approx 80$),
$k_B$ Boltzmann's constant and $T$ the absolute temperature. 
If $\delta$ is the mean distance between charges, counterion condensation is expected when $\lambda_{B} / \delta > 1$.
Considering the temperature is T=293~K and the ions are monovalent, \cite{tuszynski2004ionic} finds $\lambda_{B} = 7.13 \times 10^{-10}$~m and \cite{tusz2018} $\lambda_{B} = 13.8 \times 10^{-10}$~m for Ca$^{2+}$ at T=310~K. Considering actin \textit{filaments}, $\delta$ is estimated to be $0.25$~nm because assuming an average of 370 monomers per $\mu$m there are c. $4e$/nm. Each monomer behaves like an electrical circuit with inductive, capacitive, and resistive components. The model is based on the transmission line analogy. 

The capacitance is computed considering the charges contained in the space between two concentric cylinders, the inner with radius half the width of a monomer ($r_{actin} = 2.5$~ nm) and the outer with radius $r_{actin} + \lambda_{B}$; both cylinders are one monomer high (5.4~nm).
Thus, 
\begin{equation}
C_0 = \frac{2 \pi \epsilon l}{ln ( \frac{r_{actin} + \lambda_B}{r_{actin}} )},
\label{capacitance}
\end{equation}
where $l\approx 5.4$~nm is the length of a monomer. 

The charge on this capacitor is assumed to vary in a nonlinear way with voltage, according to the formula:
\begin{equation}
    Q_n = C_0 (V_n - b V_{n}^{2}).
\end{equation}

The inductance is computed as:
\begin{equation}
L = \frac{\mu N^2 \pi (r_{actin} + \lambda_B)^2}{l},
\label{inductance}
\end{equation}
where $\mu$ is the magnetic permeability of water and $N$ is the number of turns of the coil, that is the number of
windings of the distribution of ions around the filament. It is approximated by counting how many ions can
be lined up along the length of a monomer as $N=l/r_h$, and it is supposed that the size of a typical ion
is  $r_h \approx 3.6 \times 10^{-10}$~ m. 

The resistance is estimated considering the current between the two concentric cylinders, obtaining:
\begin{equation}
R = \frac{\rho \, ln ((r_{actin} + \lambda_B)/r_{actin})}{2 \pi l},
\label{resistance}
\end{equation}
where resistivity $\rho$ is approximately given by:
\begin{equation}
\rho = \frac{1}{\Lambda_{0}^{K^{+}} c_{K^{+}}+\Lambda_{0}^{Na^{+}} c_{Na^{+}}}.
\label{resistivity}
\end{equation}
Here, $c_{K^{+}}$ and $c_{Na^{+}}$ are the concentrations of sodium and potassium ions, which were considered in previous papers to be 0.15~M and 0.02~M, respectively; $\Lambda_{0}^{K^{+}} \approx 7.4 (\Omega m)^{-1} M^{-1}$ and $\Lambda_{0}^{Na^{+}}\approx 5.0 (\Omega m)^{-1} M^{-1}$ depend only on the type of salts. With this formula $R_1$ is computed and $R_2$ is taken as $\tilde 1/7 R_1$. Here, $R_1$ accounts for viscosity.

\begin{figure}[h]
		\includegraphics[width=1\linewidth]{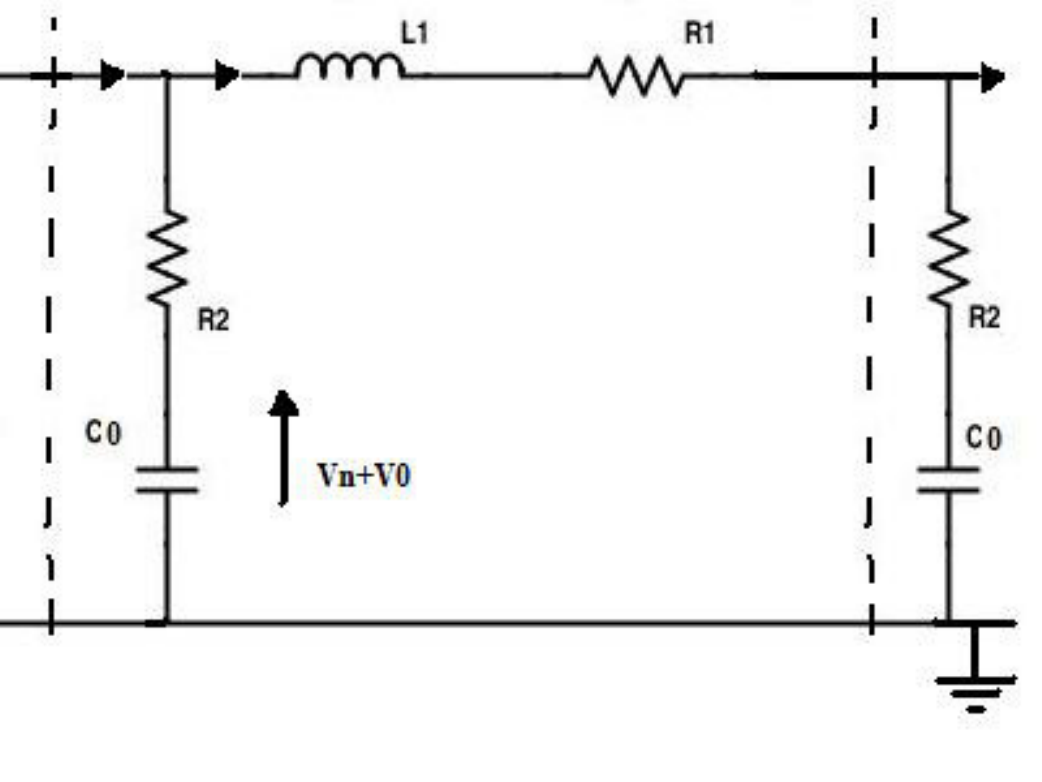}
	\caption{A circuit diagram for the $n$-th unit of an actin filament. From~\cite{tuszynski2004ionic}.}
	\label{modello}
\end{figure}

Figure~\ref{modello} illustrates the circuit schema, where an actin monomer unit in a filament is delimited by the dotted lines.

The main equation for filaments is the following, derived from \cite{tuszynski2004ionic} (see Fig.~\ref{modello} for the meaning of $R_{1}$ etc.).  

$
L C_{0} \frac{d^{2}}{dt^{2}} (V_{n} - b V_{n}^{2}) \,=
$

$
=\, V_{n+1} + V_{n-1} - 2V_{n} \,-\, R_{1}C_{0} \frac{d}{dt} (V_{n} - b V_{n}^{2})
$

$
-\, R_{2}C_{0} \{ 2 \frac{d}{dt} (V_{n} - b V_{n}^{2}) - \frac{d}{dt} (V_{n+1} - b V_{n+1}^{2})
$

\begin{equation}
-\, \frac{d}{dt} (V_{n-1} - b V_{n-1}^{2}) \}
\label{secondorder}
\end{equation}

In \cite{siccardi2016boolean} we used this equation to compute the evolution of some tens of monomers in a filament.

\section{Extension to bundle networks}

In order to extend the model to bundle networks, we must compute the suitable electrical parameters. We will consider two possibilities:
\begin{enumerate}
    \item The filament density in the bundle is so low that each filament stands at a distance greater than twice $\lambda_{B}$ from all the others. In this situation, we assume that filaments do not interact and that each one behaves as if it would not be in the bundle.
    \item The inner-bundle density is high enough that areas closer than $\lambda_{B}$ to the filaments intersect. In this situation, we will conservatively assume that the influences of the filaments' ions cancel out. 
\end{enumerate}
In case 1, we can either consider the parameters for a filament and solely multiply results by the number of filaments in the bundle or compute $C$, $L$ and $R$ using the formulas for elements connected in parallel.
In case 2, we only use the bundle radius instead of the filament one in the above formulas.

Considering a Bjerrum length $\lambda_{B} = 7.13 \times 10^{-10}$~m \cite{tuszynski2004ionic}, results for high density bundles at different bundle widths are displayed in table \ref{tabcapahigh}. Results for low density bundles made of varying filament numbers are shown in table \ref{tabcapalow}.

\begin{table}[h]
\begin{tabular}{|c|c|c|c|c|}
    \hline
Width     & 200~nm & 450~nm & 700~nm \\
\hline
$C_{0}$ in pF   &  33.8 $10^{-4}$ & 76 $10^{-4}$ & 11.8 $10^{-3}$  \\
L in pH   & 1668  & 8378  & 20227 \\
$R_{1}$ in $M\Omega$ & 0.173 & 0.077 & 0.049  \\
\hline
\end{tabular}
\caption{$C_{0}$, $L$ and $R_{1}$ for high density bundles.}
\label{tabcapahigh}
\end{table}

\begin{table}[h]
\begin{tabular}{|c|c|c|c|c|}
    \hline
Filaments     & 1 & 25 & 50 & 75\\
\hline
$C_{0}$ in pF   &  102.6 $10^{-6}$ & 4.1 $10^{-6}$ & 2 $10^{-6}$  & 1.4 $10^{-6}$ \\
L in pH   &  1.92 & 7.66 $10^{-2}$ & 3.83 $10^{-2}$  & 2.56 $10^{-2}$ \\
$R_{1}$ in $M\Omega$  & 5.7  & 0.23  & 0.11  & 0.08 \\
\hline
\end{tabular}
\caption{$C_{0}$, $L$ and $R_{1}$  for low density bundles.}
\label{tabcapalow}
\end{table}

In the following we will define equations for nodes. Equation (\ref{secondorder})  applies to elements \textbf{inside} the bundle, so we will use equation (\ref{eqnewmodel1}) instead, where $M$ is the number of edges linked to a node, and the suffix $n_k$ ranges in the set of elements of those edges that are directly connected to the node.

The term $F_n$ represents an input voltage, which is supposed to be non-zero only for some values of $n$.

$
 \frac{d^{2}}{dt^{2}} (V_{n} - b V_{n}^{2}) \,= \frac{1}{L C_{0}} \times
$

$
\times \,\{\, \sum_{k=1}^M V_{n_k} - M \times V_{n} \,+ F_n \,-\, R_{1}C_{0} \frac{d}{dt} (V_{n} - b V_{n}^{2})
$

\begin{equation}
-\, R_{2}C_{0} \{ M \times  \frac{d}{dt} (V_{n} - b V_{n}^{2}) - \sum_{k=1}^M \frac{d}{dt} (V_{n_k} - b V_{n_k}^{2}) \,\}
\label{eqnewmodel1}
\end{equation}

These equations can represent any type of element in the network. When $M=2$, they coincide with (\ref{secondorder}) and represent internal elements of a bundle. When $M=1$, they refer to a free terminal element of a bundle that is not connected to anything else. We note that in \cite{siccardi2016boolean} we used a slightly different equation for this case, namely we always kept $M=2$. The present form is more consistent with the model and its generalization. Other values of $M$ represent generic nodes.

\section{The network}

We used a stack of low-dimensional images of the three-dimensional actin network, produced in experiments on the formation of regularly spaced bundle networks from homogeneous filament solutions~\cite{huber2015formation}.  The network was chosen because it resulted from a protocol that reliably produces regularly spaced networks due to self-assembly effects~\cite{huber2015formation,glaser2016} and thus could be used in prototyping of cytoskeleton computers. From the stack of images we extracted a network description, in terms of edges and nodes, and used it as a substrate to compute the electrical behavior.
The extracted structure takes into account the main bundles in each image, with their intersections, and an estimate of bundles that can connect nodes in two adjacent images. It is not an accurate portrait of all the bundles, but it captures the main characteristics of the network.

\begin{figure}[h]
		\includegraphics[width=1\linewidth]{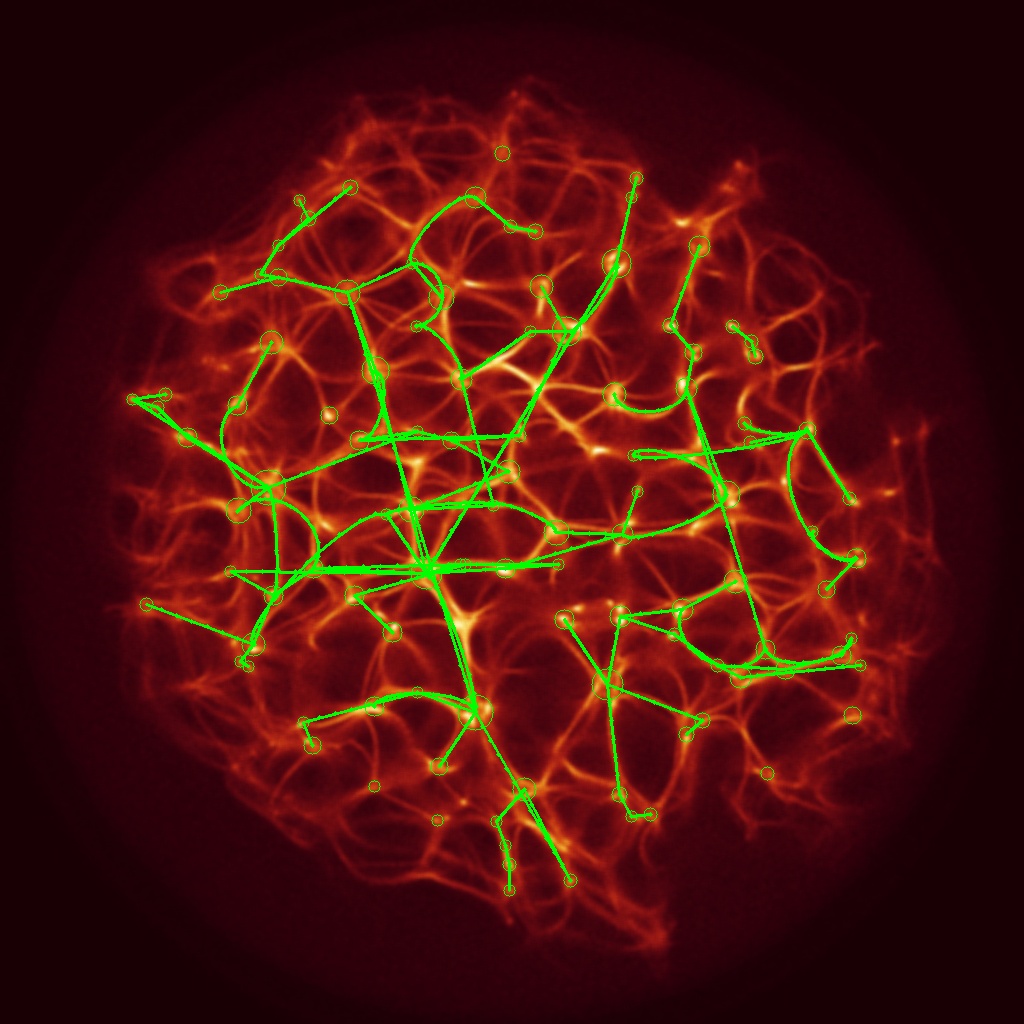}
	\caption{Two-dimensional centers and edges of a Z-slice.}
	\label{orig_04}
\end{figure}

As an illustration, in Fig.~\ref{orig_04} the green lines drawn on the original image represent the computed edges.

\begin{figure}[!tbp]
    \centering
    \includegraphics[width=0.9\linewidth]{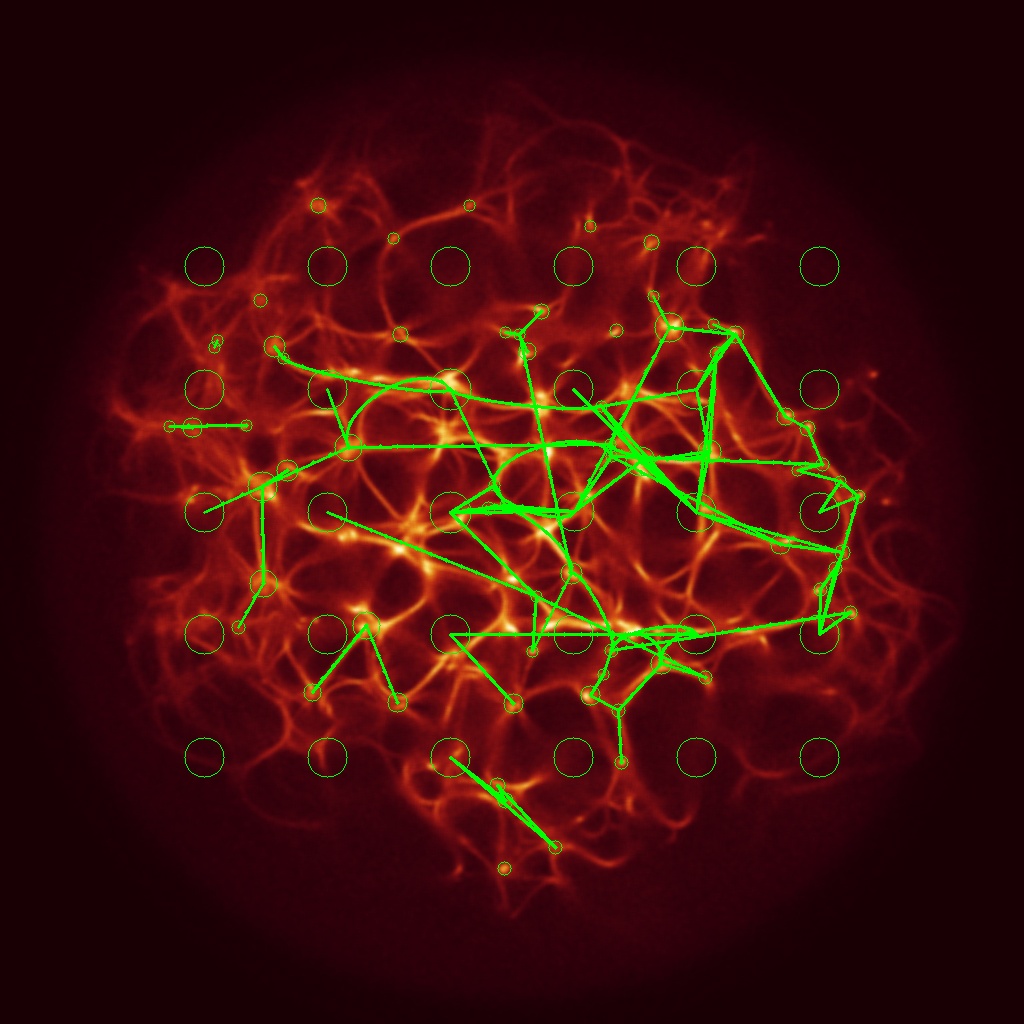}
    \caption{The grid of electrodes as it appears when the network formed on top of the supporting glass.}
    \label{orig_00_el0}
\end{figure}

In a realistic experiment, one would have to set a support with electrodes in contact with the network. In a typical configuration, we will consider a grid of 5 $\times$ 6 electrodes, for instance, on a thin glass; their diameter is $10 \mu m$ and center-to-center distances are $30 \mu m$.
We considered two situations: (1)~the network is grown in droplets sitting on the glass surface which holds the electrodes. This is actually very close to the experimental setup described in \cite{huber2012, huber2015formation}. And (2) the array of electrodes is set inside the network, i.e. in the middle of the actin droplet along its vertical axis.  This might be the case if the networks were grown around the electrode layer or if it were placed in the network later. 
Figure~\ref{orig_00_el0} is an illustration of the network grown on top of the electrode-containing glass.

The general features of the network are listed in Tab.~\ref{tab:parameters}.

\begin{table}[h]
    \centering
    \begin{tabular}{p{7cm}|l}
Parameter & Value \\ \hline
Number of nodes in the main connected graph & 2968\\
Number of edges in the main connected graph & 7583\\
Max number of nodes linked to a node & 13\\
Average number of nodes linked to a node & 5.07\\
Standard deviation of nodes linked to a node & 2.14\\
Average radius of edges in pixels & 8.48\\
Max radius of edge in pixels & 20\\
Min radius of edge in pixels & 3\\
Standard deviation of radii of edges in pixels & 2.62\\
Average edge radius, one pixel is  244.14 nm & 2.07~$\mu$m\\
Average length of edges in pixels & 70.11\\
Max length of edge in pixels & 465.40\\
Min length of edge in pixels & 4.12\\
Standard deviation of lengths of edges in pixels & 41.54\\
Average edge length, one pixel is 244.14~nm &  17.12~$\mu$m
    \end{tabular}
    \caption{Parameters of the actin network used in the modelling.}
    \label{tab:parameters}
\end{table}

\section{Preliminary results}

We used the simplest possible form for the input functions $F_n$, that is constant functions: 
\begin{equation}
    \text{for}\ 0<t<t_1 \, F_n \equiv 
    \begin{cases}
    1 & \text{if}\  n \in N_1 \\
    0 & \text{if}\  n \in N_0 \\ 
    -1 & \text{if}\ n \in N_{-1} ,
    \end{cases}
    \label{inputfunction}
\end{equation}
where $N_{1}$, $N_0$ and $N_{-1}$ are three sets of indices and $t_1$ the duration of the input stimuli, which can be equal to or less than the whole experiment time.

\begin{figure}[h]
		\includegraphics[width=1\linewidth]{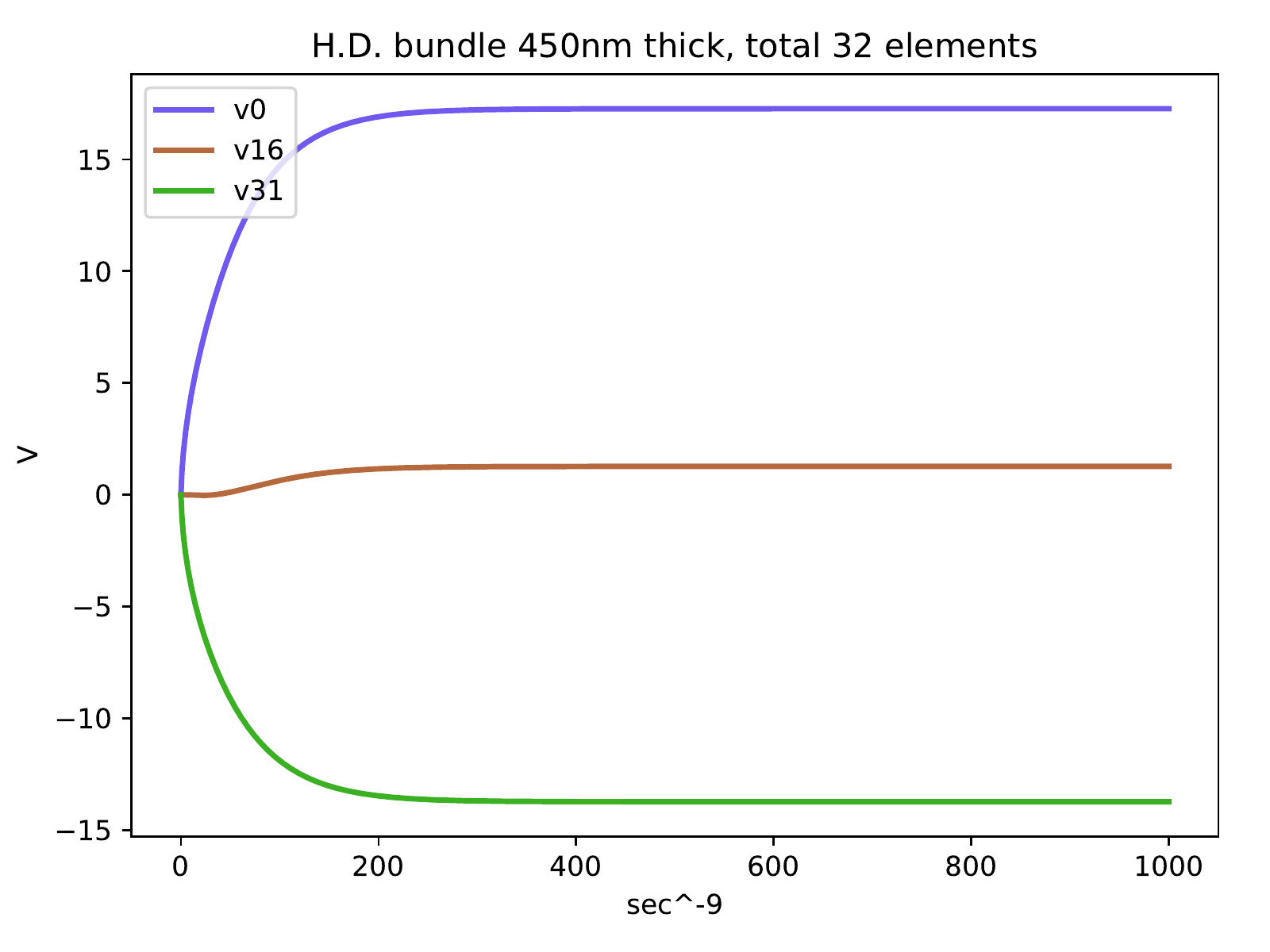}
	\caption{Evolution of some elements of an open high density (H.D.) bundle 32 elements long, with input = 1 and -1.}
	\label{stim1_1_hd}
\end{figure}

\begin{figure}[h]
		\includegraphics[width=1\linewidth]{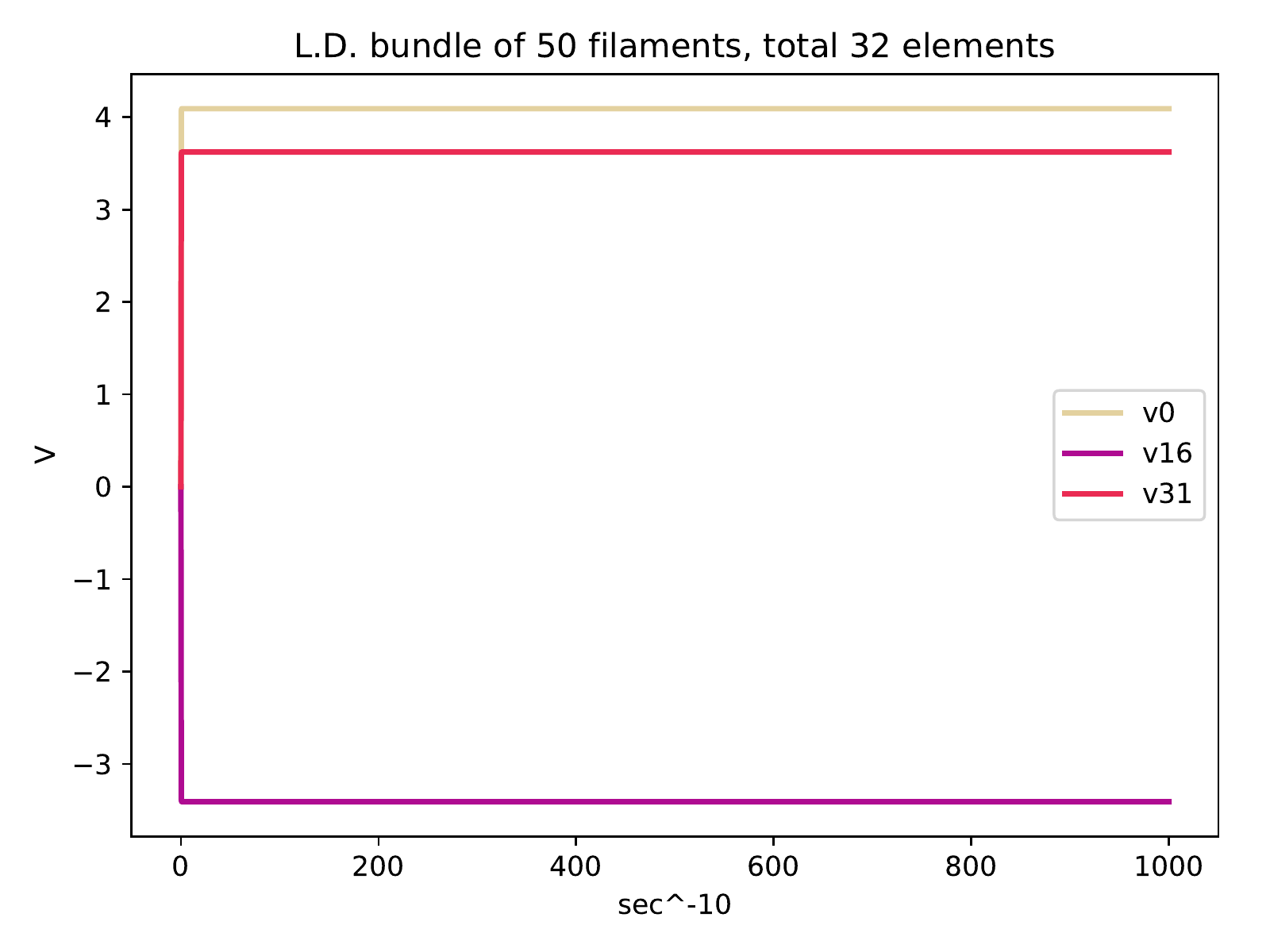}
	\caption{Evolution of some elements of a closed low density (L.D.) bundle 32 elements long, with input = 1 and -1.}
	\label{stim1_1_ldclosed}
\end{figure}

Numerical integration has been performed for open and closed bundles consisting of some tens of elements using various stimuli and electrical values. Examples can be found in Fig.~\ref{stim1_1_hd} (high density open bundle) and Fig.~\ref{stim1_1_ldclosed} (low density closed bundle).

These numerical experiments demonstrate that in all the cases considered, the solutions become constant after a transient time. Moreover, when the inputs are blocked all the solution converge to the same constant value, so that no currents can be detected.

We therefore considered constant stimuli lasting for all the experiment time and searched for constant solutions.
Input bits are defined as a pair of points of the network, so that  a +1 potential (in arbitrary units) is applied at one of them and -1 at the other to encode a value 1 of the bit; when no potential is applied, the bit value is zero.

Analogously, we chose pairs of points and measured the difference of their potential to read an output bit. A suitable threshold has been defined to distinguish the 1 and 0 values. 

\section{Results}

\subsection{Ideal electrodes}

In this section we consider ideal electrodes that (1)~can be placed in any point on the network surface or inside it and (2)~are so small that they would be in contact with one element of a single bundle only. Moreover we use a slightly idealized network of spherical shape. 

\subsubsection{Boolean gates}

We randomly chose 8 sites in the network and considered them as 4 pairs to represent 4 input bits.

Then we applied in turn all the possible input states from (0000) to (1111) and solved the system (\ref{eqnewmodel1}). It reduces to a linear algebraic structure and  simplifies finding the values of the potential in the \textbf{nodes}.
Then we checked, for all the sets of input states that correspond to a logical input, which output bits correspond to the expected results for a gate.

For instance, to find the {\sc not} gates we considered input state sets ((0000),(0001)), ((1000),(1001)) etc.; then we looked for all the output bits that are 1 for the first state and 0 for the second of one of the input sets. 

The same procedure was used to find {\sc or}, {\sc and} and {\sc xor} gates. We used three values for the output threshold: 2, 1, and 0.5.

The results of three runs are shown in Tabs. \ref{tabnotgates}, \ref{taborgates}, \ref{tabandgates} and \ref{tabxorgates} revealing that once an input position is chosen, it is possible to find a suitable number of edges that behave as output for the main gate types.

\begin{table}[h]
\begin{tabular}{|c|c|c|c|c|}
Run     & Thresh. 2 & Thresh. 1 & Thresh. 0.5 \\
    \hline
1     & 8266 & 8944 & 8409\\
2     & 3688 & 4660 & 4682\\
3     & 5730 & 7043 & 7455
\end{tabular}
\caption{Number of possible {\sc not} gates}
\label{tabnotgates}
\end{table}

\begin{table}[h]
\begin{tabular}{|c|c|c|c|c|}
Run     & Thresh. 2 & Thresh. 1 & Thresh. 0.5 \\
    \hline
1     & 4385 & 8191 & 12494\\
2     & 6360 & 8188 & 11336\\
3     & 5835 & 8260 & 11063
\end{tabular}
\caption{Number of possible {\sc or} gates}
\label{taborgates}
\end{table}

\begin{table}[h]
\begin{tabular}{|c|c|c|c|c|}
Run     & Thresh. 2 & Thresh. 1 & Thresh. 0.5 \\
    \hline
1     & 3600 & 3562 & 2577\\
2     & 4506 & 43842 & 3119\\
3     & 4954 & 5076 & 3726
\end{tabular}
\caption{Number of possible {\sc and} gates}
\label{tabandgates}
\end{table}

\begin{table}[h]
\begin{tabular}{|c|c|c|c|c|}
Run     & Thresh. 2 & Thresh. 1 & Thresh. 0.5 \\
    \hline
1     & 1543 & 2155 & 3749\\
2     & 584 & 986 & 1799\\
3     & 1009 & 1499 & 3083
\end{tabular}
\caption{Number of possible {\sc xor} gates}
\label{tabxorgates}
\end{table}

\subsubsection{Time estimates}

We also computed the time that the network would need to converge to the constant solutions, taking into account the time needed for an element to discharge. As a first estimate, we used the value $R_1C_0$, that is the discharge time of a pure RC circuit. Using the parameters for a single filament (or for low density bundles made of independent filaments), we got $2.248 \cdot 10^{-3}$~sec to travel the 3843876 elements of the whole network. Parameters for a high density network, adjusted for the estimated width of each bundle, give a time of $2.25 \cdot 10^{-3}$~sec. In both cases, the velocity is of the order of 4.7~m/sec, two orders of magnitude larger than the estimate found in \cite{tusz2018} with a different model (pure RC), but in the range estimated in \cite{tuszynski2004ionic} using the present one.

\subsection{Realistic electrodes}

In this section we consider electrodes that could be actually available, with their supporting glass.
Moreover, we use the real network dimensions (the confocal images are $250 \mu m \times 250 \mu m$ and they are spaced $110 \mu m$ in depth).

We considered both the case with the network being on top of the glass holding the electrodes, and the case when the electrodes are placed inside the network, along the middle plane of the confocal image stack.

\subsubsection{Boolean gates}

In the case of the network on top of the glass, we randomly chose 8 electrodes and considered them as 4 pairs to represent 4 input bits.

We applied in turn all the possible input states from (0000) to (1111) and solved the system (\ref{eqnewmodel1}). Then we computed the potential differences for all the pairs of electrodes that were not used as input and applied a suitable threshold to distinguish 0 and 1 bits. The threshold we used was the median of the differences.

As 10 electrodes out of the 18 connected to the network were not used as input, we had 45 potential output bits.
We found that, considering all the possible input and output bits, we have 101 {\sc not} gates, 113 {\sc or} gates, 46 {\sc and} gates and 13 {\sc xor} gates. 

It must be noted that the same pair of output electrodes may have been counted many times in these figures. For instance, the potential difference of electrodes between  
46th and 32nd electrode (electrodes in row 4 column 6 and in row 3 column 2), were considered a possible {\sc not} gate for all the cases listed in table \ref{tab4632}.

\begin{table}[h]
\begin{tabular}{|c|c|c|c|c|}
Input state     & Output value & {\sc not} on bit \\
    \hline
1100     & 1 & 4\\
1101     & 0 & \\
    \hline
1010     & 1 & 4\\
1011     & 0 & \\
    \hline
1000     & 1 & 4\\
1001     & 0 & \\
    \hline
0110     & 1 & 4\\
0111     & 0 & \\
    \hline
0100     & 1 & 4\\
0101     & 0 & \\
    \hline
0111     & 1 & 2\\
0011     & 0 & \\
    \hline
0101     & 1 & 2\\
0001     & 0 & \\
    \hline
1011     & 1 & 1\\
0011     & 0 & \\
    \hline
1001     & 1 & 1\\
0001     & 0 & \\
    \hline
\end{tabular}
\caption{Possible {\sc not}  with a single edge}
\label{tab4632}
\end{table}

In the case of the network with electrodes placed in the interior, we randomly chose 12 electrodes and considered them as 6 couples to represent 6 input bits.

We applied in turn all the possible input states from (000000) to (111111) and solved the system (\ref{eqnewmodel1}). Then we computed the potential differences for all the pairs of electrodes that were not used as input and applied a suitable threshold to distinguish 0 and 1 bits. The threshold we used was the median of the differences.

As 15 electrodes out of the 27 connected to the network were not used as input, we had 105 potential output bits.
We found that, considering all the possible input and output bits, we have 1885 {\sc not} gates, 1279 {\sc or} gates, 783 {\sc and} gates and 467 {\sc xor} gates. 

\section{Finite state machine}

The actin network implements  a mapping from $\{0, 1\}^k$ to $\{0, 1\}^k$, where $k$ is a number of input bits, represented by potential difference in pairs of electrodes, as described above. Thus, the network can be considered as an automaton or a finite state machine, ${\mathcal A}_k = \langle \{ 0, 1 \}, C, k, f \rangle$. The behaviour of the automaton is governed by the function $f: \{0, 1\}^k \rightarrow  \{0, 1\}^k$, $k \in \mathbf{Z}_+$. The structure of the mapping $f$ is determined by exact configuration of electrodes $C \in \mathbf{R}^3$ and geometry of the AF bundle network. 

\subsubsection{Using two values of $k=4$ and $k=6$.}

The machine ${\mathcal A}_4$ represents the actin network placed onto an array of electrodes. In this case, having at our disposal 45 potential output bits, the number of combinations of 4 of them is 148995. We therefore limited the study at the output positions that assume a 1 value more than 6 and less than 11 times for the 16 input states. In this way we found 11 output bits and computed the state transitions for the 330 machines that one can obtain choosing 4 out of them, $k=4$.

The machine ${\mathcal A}_6$ represents the actin network where the array of electrodes is inside the network. 
In this case, having at our disposal 105 potential output bits, the number of combinations of 6 of them is quite large. We therefore limited the study at the output positions that assume a 1 value 32 times for the 64 input states. In this way we found again 11 output bits and computed the state transitions for the 462 machines that one can obtain choosing 6 out of them, $k=6$. 

We derived  structures of functions $f_4$ and $f_6$, governing behavior of automata ${\mathcal A}_4$ and ${\mathcal A}_6$, as follows.
There is potentially an infinite number of electrode configurations from $\mathbf{R}^3$. Therefore, we selected 330 and 462 configurations $C$ for machines $\mathcal{A}_4$ and $\mathcal{A}_6$, respectively, and calculated the frequencies of connections of input to output states, obtaining two probabilistic state machines  $= \langle \{ 0, 1 \},  p, k, f \rangle$, where $p: {\{0, 1\}^k}^{\{ 0,1 \}} \rightarrow [0,1]$, the $p$ assigns a probability to each mapping from $\{0,1\}^k$ to $\{ 0, 1 \}$. Thus, a state transition of $\mathcal{A}_k$ is a directed weight graph, where weight represents a probability of the transition between states of $\mathcal{A}_k$ corresponding to nodes of the graph.  The weighted graph can be converted to a non-weighted directed graph  by removing all edges with weight less than a given  threshold $\theta$. Let us perform trimming for several thresholds with 0.1 increment.

\begin{figure*}[!tbp]
    \centering
    \subfigure[]{\includegraphics[width=0.49\textwidth]{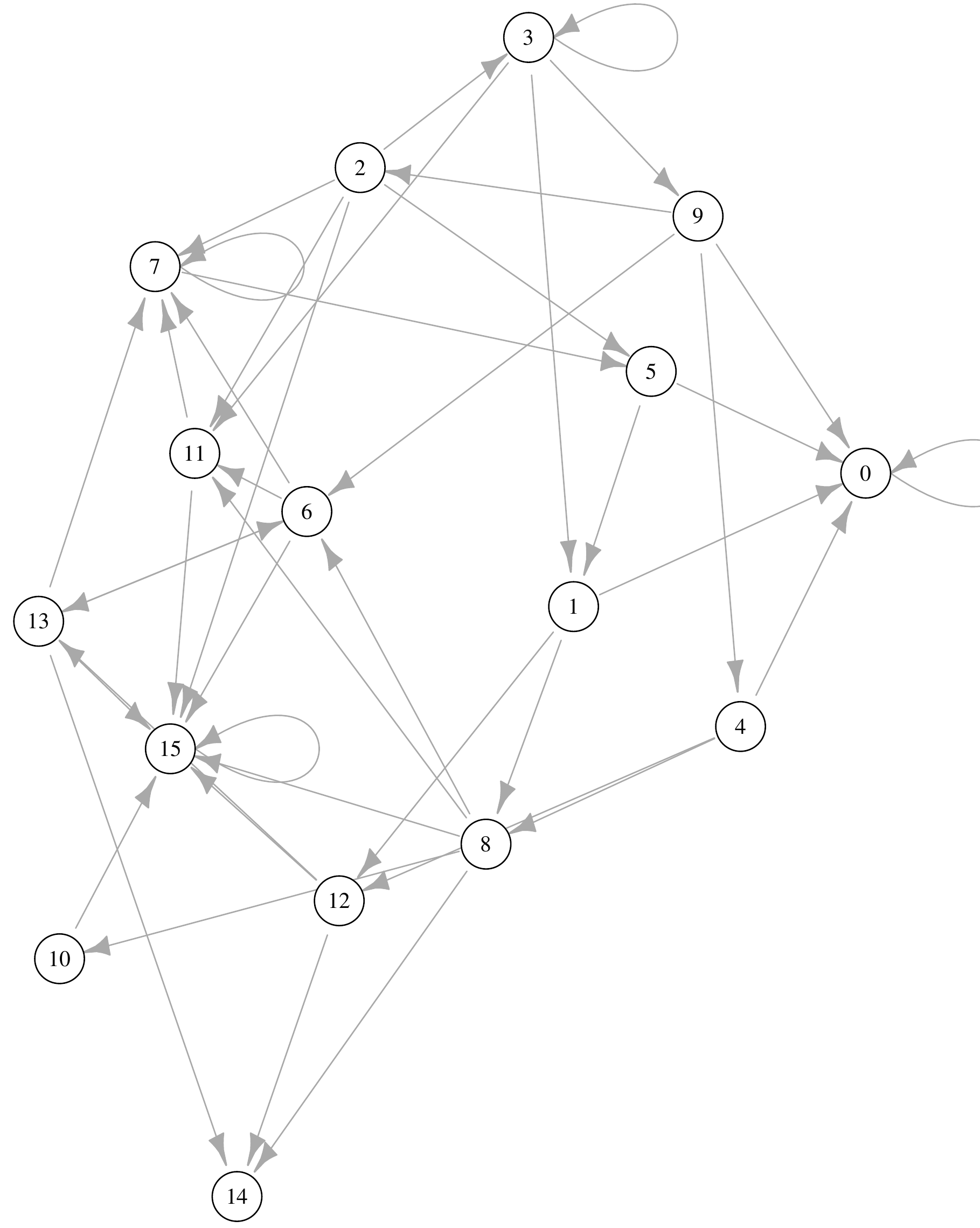}}
\subfigure[]{\includegraphics[width=0.49\textwidth]{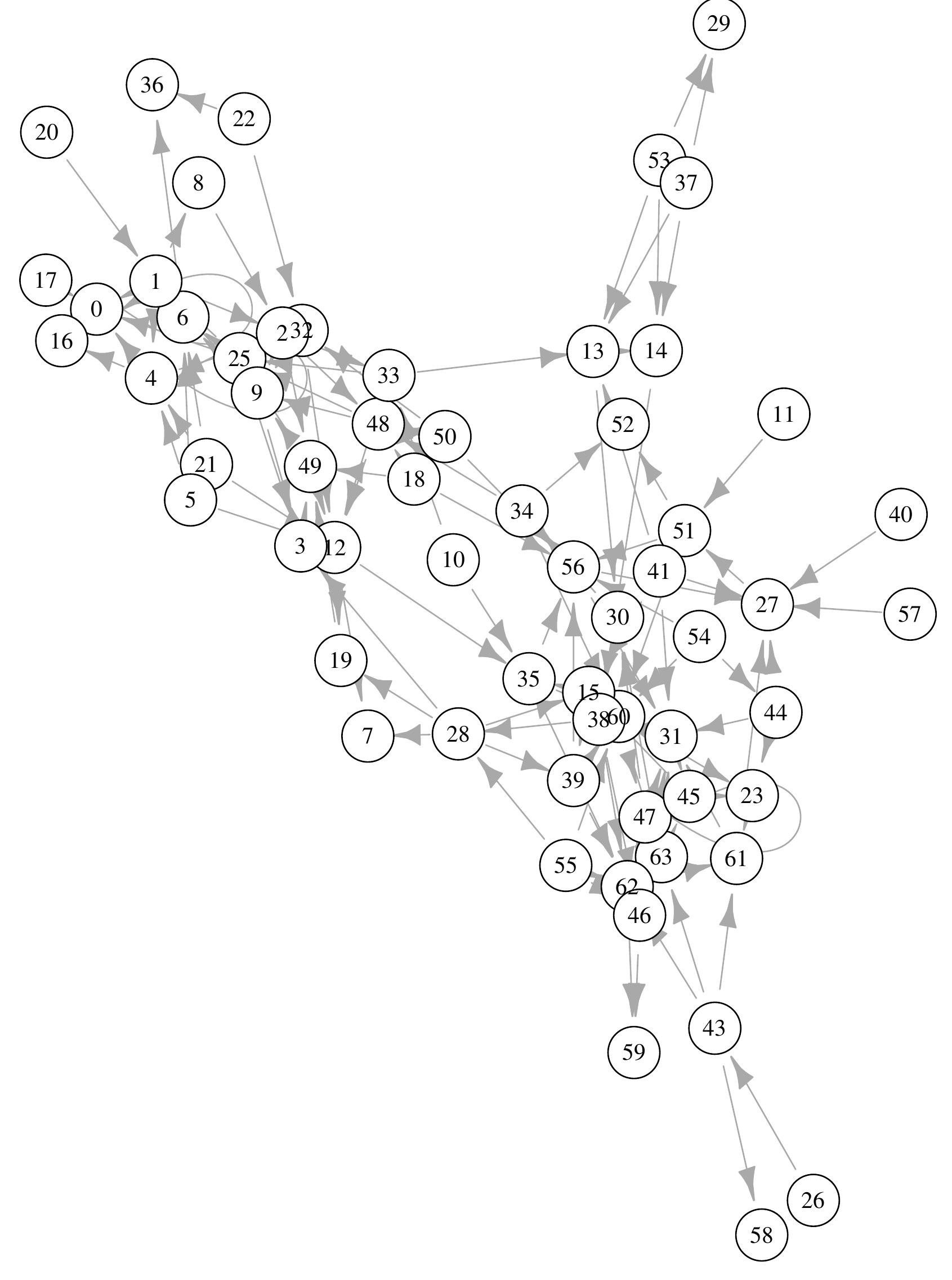}}
\label{threshold01}
\caption{State transitions graphs for (a)~$\mathcal{A}_4$ and (b)~$\mathcal{A}_6$, trimming threshold is $\theta=0.1$.  Nodes are labelled by digital representation of 4-bit~(a) and 6-bit~(b) states.}
\end{figure*}

The graph remains connected for $\theta$ till 0.1 (Fig.~\ref{threshold01}).  The graph for $\mathcal{A}_4$ is characterising for having no unreachable nodes and several absorbing states (Fig.~\ref{threshold01})a) while the graph for $\mathcal{A}_6$ has a number of unreachable nodes (Garden-of-Eden states) and less, than $\mathcal{A}_4$, absorbing states (Fig.~\ref{threshold01})b).

\begin{figure*}[!tbp]
    \centering
    \subfigure[]{\includegraphics[width=0.49\textwidth]{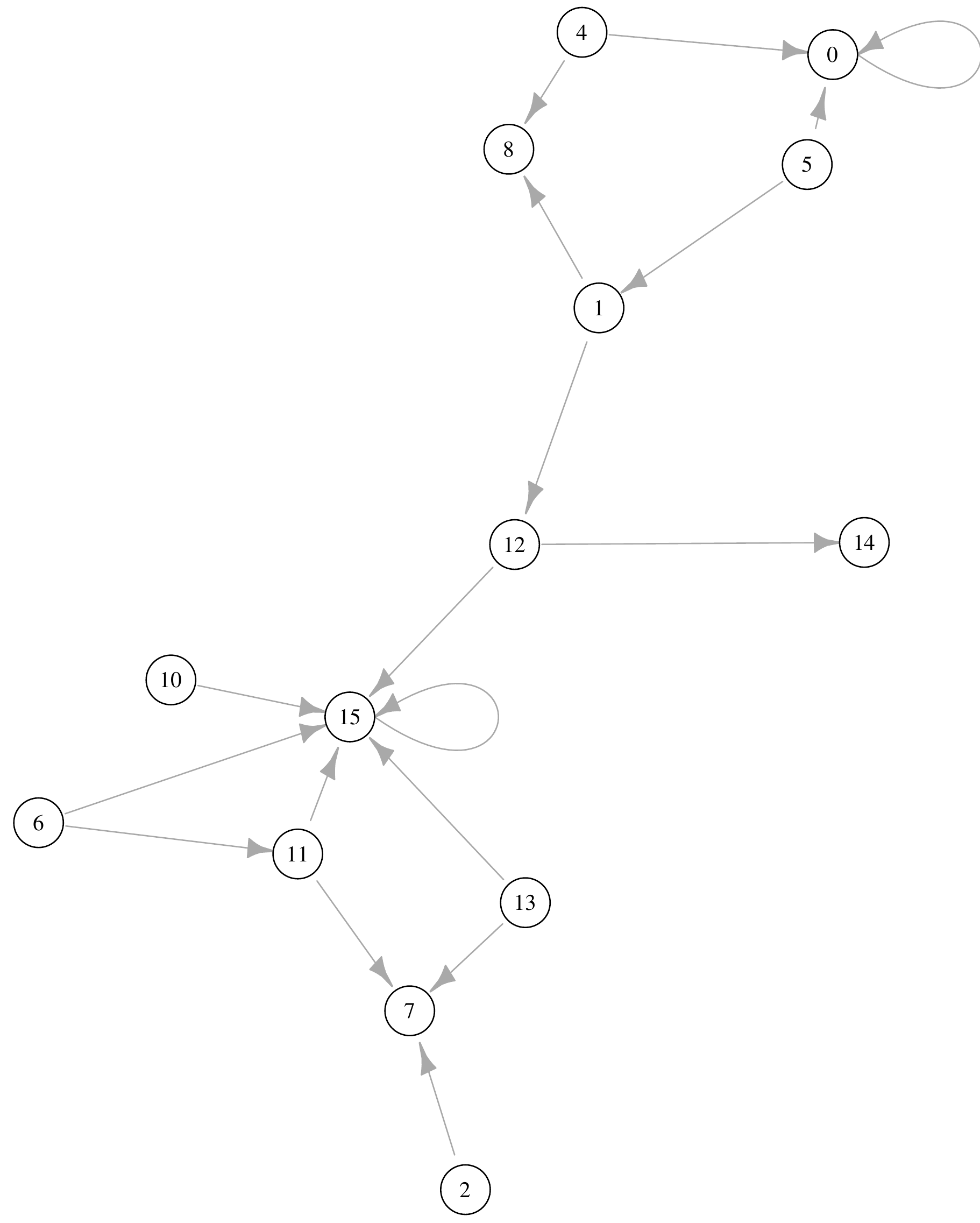}}
\subfigure[]{\includegraphics[width=0.49\textwidth]{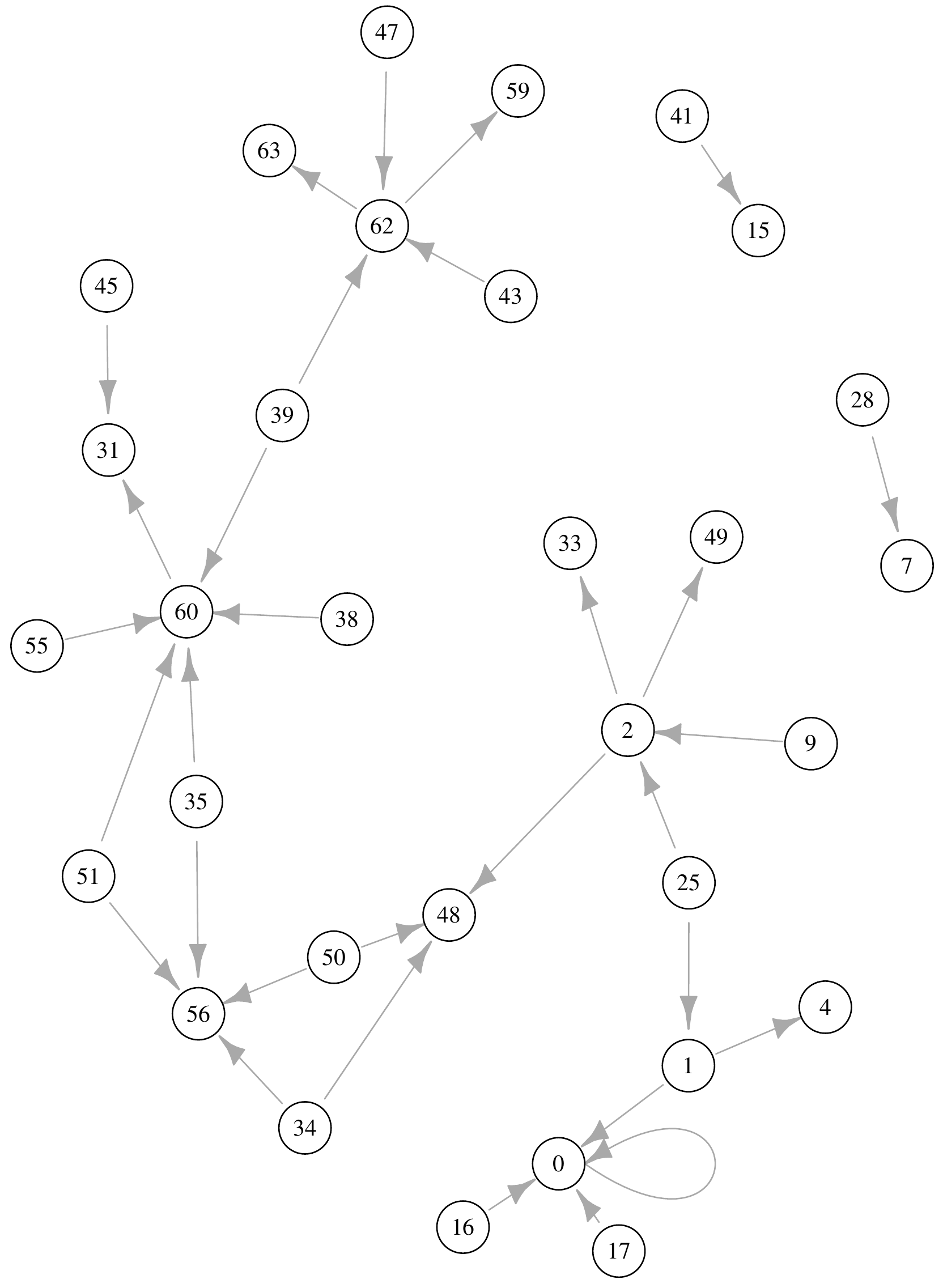}}
\label{threshold02}
\caption{State transitions graphs for (a)~$\mathcal{A}_4$ and (b)~$\mathcal{A}_6$, trimming threshold is $\theta=0.2$.  Nodes are labelled by digital representation of 4-bit~(a) and 6-bit~(b) states, respectively.}
\end{figure*}

The state transition graph of $\mathcal{A}_6$ becomes disconnected for $\theta=0.2$ (Fig.~\ref{threshold02}b) and the graph of $\mathcal{A}_6$ remains connected (Fig.~\ref{threshold02}b).  

\begin{figure*}[!tbp]
    \centering
    \subfigure[]{\includegraphics[width=0.49\textwidth]{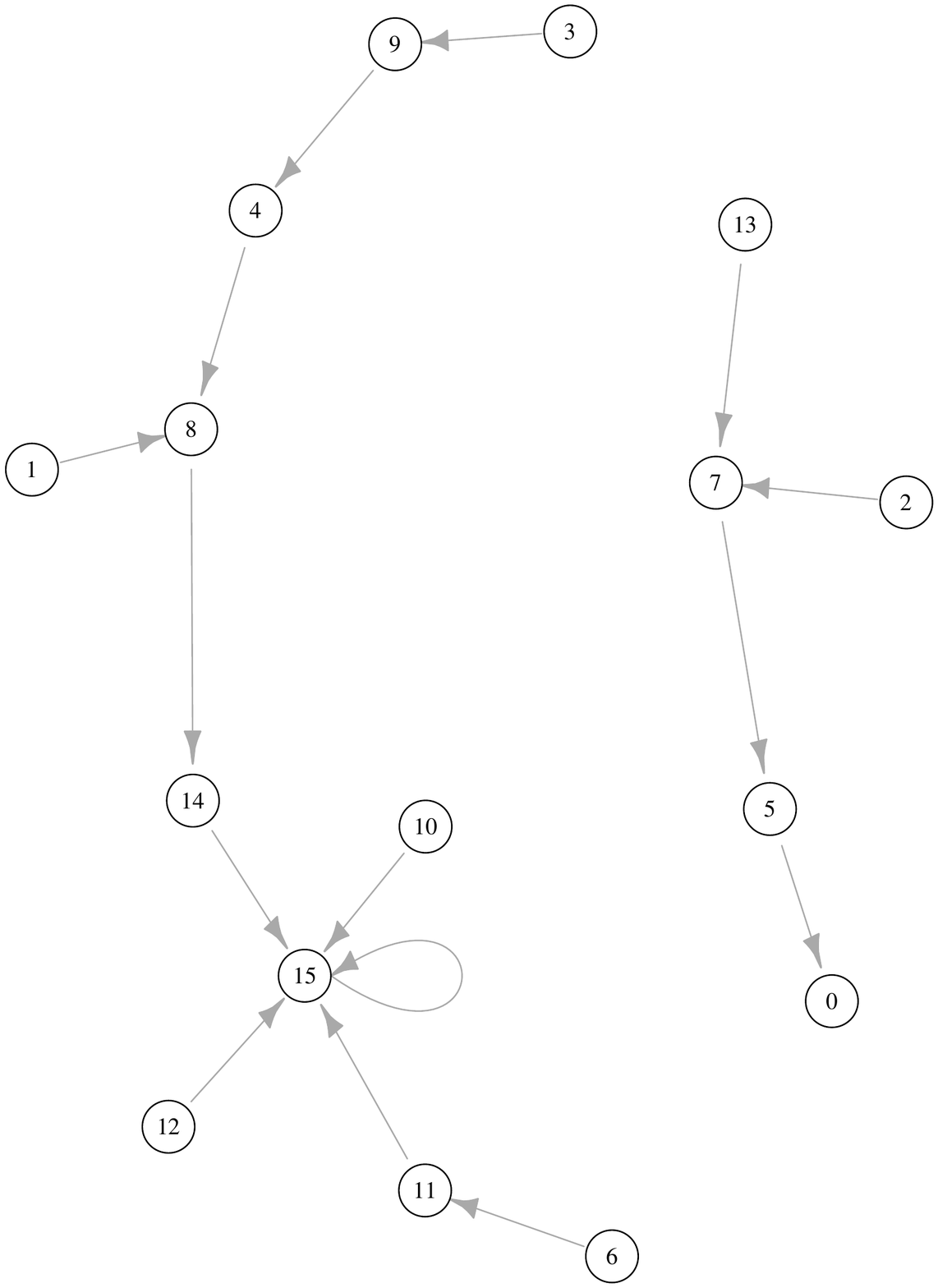}}
    \subfigure[]{\includegraphics[width=0.49\textwidth]{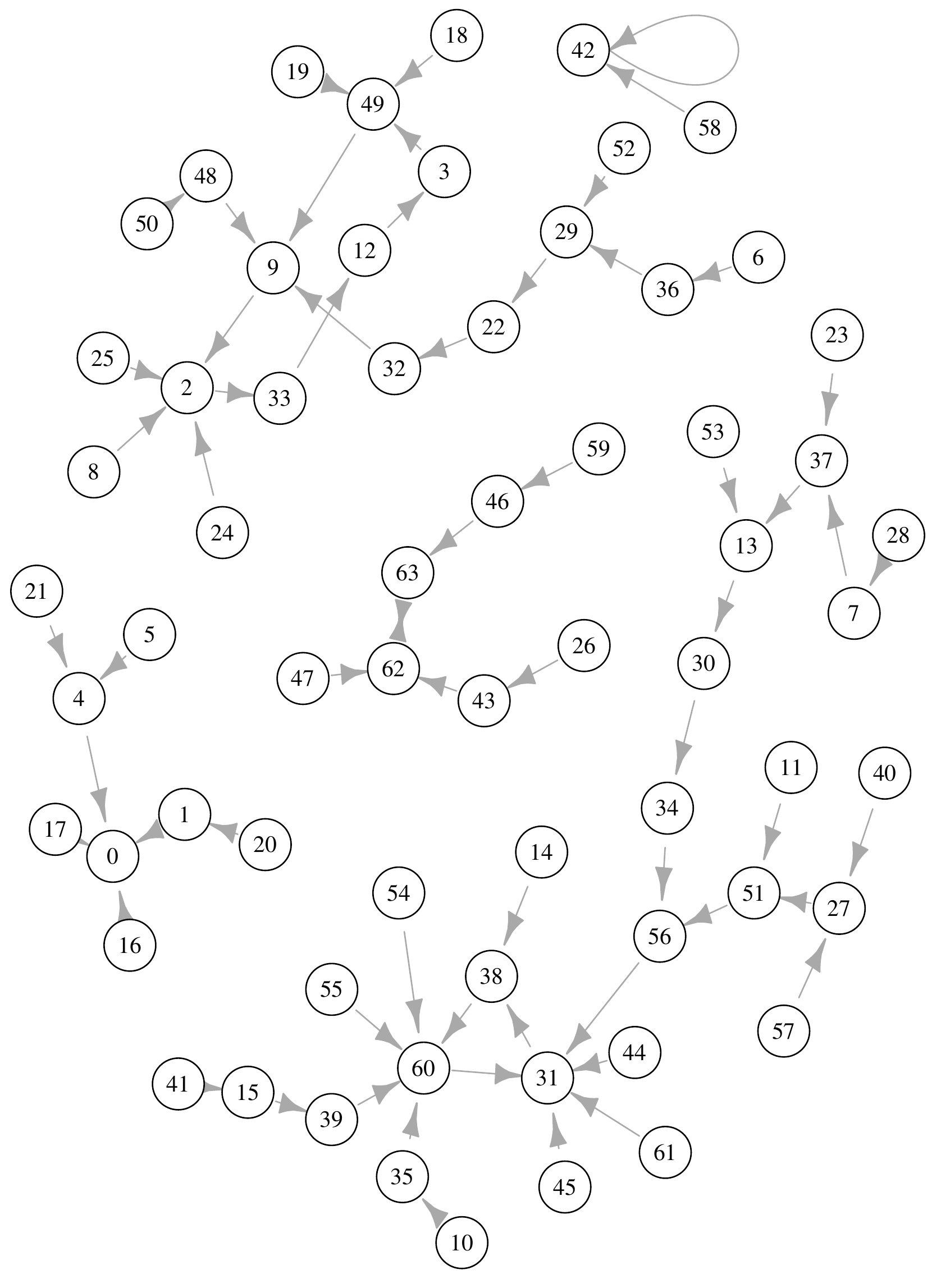}}
    \caption{Graphs representing most likely transitions $G_4$~(a) and $G_6$~(b) of $\mathcal{A}_4$~(a) and $\mathcal{A}_6$~(b).}
    \label{fig:deterministicmachines}
\end{figure*}

Another way of converting weighted, probabilistic, state transition graphs into non-weighted graphs is by selecting for each node $x$ a successor $y$ such that the weight of the arc $(xy)$ is the highest among all arcs outgoing from $x$. These graphs $G_4$ and $G_6$ of most likely transitions are shown in Fig.~\ref{fig:deterministicmachines}. The graph $G_4$ (Fig.~\ref{fig:deterministicmachines}a) has two disconnected sub-graphs, 8 Garden-of-Eden states and two absorbing states corresponding to $(1111)$ and $(0000)$; the graph has no cycles. The graph $G_6$ has 5 disconnected sub-graphs (Fig.~\ref{fig:deterministicmachines}a). Two of them have only absorbing states, corresponding to $(00000)$ and $(101010)$, and no cycles. Three of the sub-graphs do not have an absorbing state but have cycles: 
$ (111110) \rightarrow (111111) \rightarrow (111110)$, 
$(001111) \rightarrow (001111) \rightarrow (100110) \rightarrow (111100) \rightarrow (001111)$ and 
$
(000011) \rightarrow
(110001) \rightarrow
(001001) \rightarrow
(000010) \rightarrow
(100001) \rightarrow
(001100) \rightarrow
(000011) 
$.
 
\section{Discussion}

By using a physical model of ionic currents on non-linear transmission network we demonstrated how a computation of Boolean functions can be implemented on actin networks and what type of distribution of Boolean gates can be obtained. In the model, we employed a geometry of the three-dimensional actin bundle network derived from experimental laboratory data. Our results might act as a feasibility study for future experimental laboratory prototypes of cytoskeleton computing devices. We have also derived finite state machines realizable on the actin networks. The importance of the machines is two-folded. First, their state transition graphs might act as unique fingerprints of actin networks formed at the different experimental or physiological conditions. Second, the structure of the machines could advance studies in computational power  of actin networks in the context of formal language recognition.

The following issues could be addressed in the future. We did not consider the currents generated by ions flowing in the liquid medium containing the network. These could produce some amount of noise. The model could be improved in this aspect. We did not take into account the fact that differences of electrical potential along the bundles could give rise to local patterns of ion concentrations, and these, in turn could change the bundle resistance $R_{1}$. This could be a retrofit mechanism that we propose to study in future work.

It should also be noted that similar implementations of the model involving MTs instead of AFs are possible with minor modifications. Random arrangements of MTs in buffer solutions have been analyzed experimentally regarding their conductive and capacitive properties~\cite{kalra2019behavior,guzman2019tubulin,santelices2017response}. It was observed that MTs measurably increase the solution’s conductance compared to free tubulin at lower ionic concentrations while the opposite is true at high ionic concentrations. This effect can be explained using the Debye-Hueckel model as due to the formation of a counter-ionic layer whose thickness is concentration and temperature dependent according to the Debye length formula. At high ionic concentrations MTs act as low-resistance cables while at low ionic concentrations their contribution to impedance is mainly capacitive. At the peak value, the intrinsic conductivity of MTs has been found to be two orders of magnitude greater than that of the buffer solution. On the other hand, at low protein concentration, free tubulin dimers decrease solution’s conductance, and it was modelled as being due to tubulin attracting ionic charges and lowering their mobility. Both tubulin and MTs were found to increase capacitance of buffer solutions, due to their formation of ionic double layers. Consequently, MT networks exhibit fascinating electrical properties, which change as a function of ionic concentration and pH providing an opportunity to additionally control the functional characteristics of the networks assembled from MTs. We intend to derive a complete model based on MT networks in a future study.

\bibliographystyle{plain}
\bibliography{references}
\end{document}